\documentclass[prd,showpacs,reprint]{revtex4-1}

\usepackage{epsfig,amsmath,amssymb}
\usepackage[outdir=./]{epstopdf}

\begin{document}
	
\title{Foldy-Wouthuysen transformation of the generalised Dirac Hamiltonian in a gravitational-wave background} 

\author{James Q. Quach} 
\email{quach.james@gmail.com}
\affiliation{Institute for Solid State Physics, University of Tokyo, Kashiwa, Chiba 277-8581, Japan}

\begin{abstract}
Goncalves \textit{et al.}~\cite{goncalves07} derived a non-relativistic limit of the generalised Dirac Hamiltonian in the presence of a gravitational wave, using the \textit{exact} Foldy-Wouthuysen transformation. This gave rise to the intriguing notion that spin-precession may occur even in the absence of a magnetic field. We argue that this effect is not physical as it is the result of a gauge-variant term that was an artefact of a flawed application of the exact Foldy-Wouthuysen transformation. In this paper we derive the correct non-relativistic limit of the generalised Dirac Hamiltonian in the presence of a gravitational wave, using both the exact and \textit{standard} Foldy-Wouthuysen transformation. We show that both transformations consistently produce a Hamiltonian where all terms are gauge-invariant. Unfortunately however, we also show that this means the novel spin-precession effect does not exist.
\end{abstract}

̀\pacs{04.30.-w,04.62.+v,03.65.Pm,04.80.Nn}

\maketitle

\section{Introduction}

Gravitational waves (GWs) are one of the major predictions of general relativity that has yet to be directly observed. Light interferometry, such as that used in LIGO and VIRGO, currently stands as the most promising means by which to detect GWs. More recently however, matter-wave interferometry has been proposed as a more sensitive way to detect GWs because of the increased gravitational interaction from the massive particles~\cite{scully93,godun01,chiao04,delva06,foffa06,dimopoulos08,gao11,graham13}. In light of this, it is important to understand the quantum interaction of massive particles with GWs in the non-relativistic limit.
Goncalves \textit{et al.}~\cite{goncalves07} was the first to investigate the Dirac Hamiltonian in the presence of an electromagnetic (EM) gauge field and GWs, in the non-relativistic limit. In their work, they came to the intriguing conclusion that in the presence of the GW, the particle's spin may precess even in the absence of a magnetic field, which they propose could be the basis for a new type of GW detector. In this paper we show that this precession cannot be physical and is the result of a miscalculation. 

A systematic scheme by which to write down the non-relativistic limit of the Dirac Hamiltonian with relativistic correction terms is provided by the Foldy-Wouthuysen (FW) transformation~\cite{foldy78}. The FW transformation is a unitary transformation which separates the upper and lower spinor components. In the FW representation, the Hamiltonian and all operators are block-diagonal (diagonal in two spinors). It is an extremely useful representation because the relations between the operators in the FW representation are similar to those between the respective classical quantities. Two popular variants of the FW transformation exists: the \textit{chiral} or \textit{exact} FW (EFW) transformation~\cite{eriksen58,nikitin98,obukhov01,jentschura14} and the \textit{standard} FW (SFW) transformation~\cite{foldy78}. 

The EFW transformation is so-called because under certain anti-commutative conditions, the scheme leads to an exact expression containing only even terms: \textit{even} terms do not mix the upper and lower spinor components, \textit{odd} terms do. The EFW transformation has the advantage that its calculation is relatively simpler than the SFW transformation, however the EFW may give rise to parity-variant terms which are difficult to physically interpret~\cite{nicolaevici02,obukhov02a,silenko05,jentschura14}. The SFW transformation involves multi-commutator expansions of a unitary transformation, which eliminates the odd contributions to a desired level of accuracy. As it is an iterative scheme, its calculation can be considerably more difficult than the EFW scheme. The SFW transformation however consistently yields terms that are amenable to physical interpretation. 

To arrive at their conclusion, Goncalves \textit{et al}. applied the EFW transformation to the Dirac Hamiltonian to produce a Hamiltonian which contains a gauge-variant term. It is this gauge-variant term that gives rise to the novel spin-precession effect. We contend that this term is an artefact of a flawed application of the EFW transformation. In this work we will derive the correct non-relativistic limit of the Dirac Hamiltonian in the background of an EM gauge field and GWs, first using the EFW transformation, then as a further check, using the SFW transformation. We show that both these methods produce a gauge-invariant Hamiltonian. As a result we will show that there is no spin-precession when there is no magnetic field, contrary to the conclusions of Goncalves \textit{et al}.

In Sec.~\ref{sec:dirac} we write down the generalised Dirac Hamiltonian with a GW metric. In Sec.~\ref{sec:efw} we briefly review the work of Goncalves \textit{et al.} before deriving the correct non-relativistic limit of the generalised Dirac Hamiltonian with GWs under the EFW transformation. Sec.~\ref{sec:sfw} derives the same Hamiltonian under the SFW transformation, thereby confirming the consistency of the Hamiltonian. From this Hamiltonian we will also derive the spin equation of motion.

\section{Dirac Hamiltonian with Gravitational Wave Metric}
\label{sec:dirac}

The wave function $\psi$ of a spin-1/2 particle of rest mass $m$ and charge $e$ in an EM and gravitational field obeys the curved spacetime Dirac equation (SI units),
\begin{equation}
	i\hbar\gamma_ae_\mu^a(\partial_\mu-\Gamma_\mu-\frac{ie}{\hbar}A_\mu)\psi=mc\psi~.
\label{dirac}
\end{equation}
The spacetime metric $g_{\mu\nu}$ can be related at every point to a tangent Minkowski space $\eta_{ab}$ via tetrads $e_\mu^a$, $g_{\mu\nu}=e_\mu^a e_\nu^b\eta_{ab}$. The tetrads obey the orthogonality conditions $e_\mu^a e_a^\nu=\delta_\mu^\nu,e_\mu^a e_b^\mu=\delta_b^a$. We use the convention that Latin indices represent components in the tetrad frame. The spinorial affine connection $\Gamma_\mu=\frac{i}{4} e_\nu^a (\partial_\mu e^{\nu b}+\Gamma_{\mu \sigma}^\nu e^{\sigma b}) \sigma_{ab}$, where $\Gamma_{\mu\sigma}^\nu$ is the affine connection and $\sigma_{ab}\equiv\frac{i}{2}[\gamma_a,\gamma_b]$ are the generators of the Lorentz group. $\gamma_a$ are gamma matrices defining the Clifford algebra $\{\gamma_a,\gamma_b\}=-2\eta_{ab}$, with spacetime metric signature ($-,+,+,+$). $A_\mu$ is the EM four-vector potential. We use the Einstein summation convention where repeated indices ($\mu,\nu,\sigma,a,b=\{0,1,2,3\}$) are summed.

The metric for one of the polarisation states of the linear plane GW is,
\begin{equation}
	ds^2=-c^2dt^2+dx^2+(1-2f)dy^2+(1+2f)dz^2~.
\label{gw_metric}
\end{equation}
where $f=f(t-x)$ is a function which describes a wave propagating in the $x$-direction. Under this metric Eq.~(\ref{gw_metric}) can be written in the familiar Schr\"{o}dinger picture $i\hbar\partial_t \psi=H\psi$, where ($\boldsymbol{\alpha}\equiv\gamma^0\boldsymbol{\gamma},\beta\equiv\gamma^0,\boldsymbol{p}\equiv-i\hbar\nabla$, and $i,..,n=\{1,2,3\}$)~\cite{goncalves07},
\begin{equation}
	H=c\alpha^i(p_i-eA_i)-cf\alpha^2(p_2-eA_2)+cf\alpha^3(p_3-eA_3)+\beta mc^2.
\label{H_GW}
\end{equation}

\section{Exact Foldy-Wouthuysen Transformation}
\label{sec:efw}

Central to the EFW transformation is the property that when $H$ anti-commutes with $J\equiv i\gamma^5 \beta$, $\{H,J\}=0$, under the unitary transformation $U=U_2 U_1$, where ($\Lambda\equiv H/\sqrt{H^2}$) 
\begin{equation}
	U_1=\frac{1}{\sqrt{2}} (1+J\Lambda),\quad\quad U_2=\frac{1}{\sqrt{2}} (1+\beta J)~,
\end{equation}
the transformed Hamiltonian is even, 
\begin{equation}
\begin{split}
	UHU^+=&\frac12\beta(\sqrt{H^2}+\beta\sqrt{H^2}\beta)+\frac12(\sqrt{H^2}-\beta\sqrt{H^2}\beta)J\\
		=&\{\sqrt{H^2}\}_\text{even}\beta+\{\sqrt{H^2}\}_\text{odd}J~.
\end{split}
\label{UHU}
\end{equation}

In Eq.~(\ref{UHU}) we have made use of the fact that the even and odd components of any operator $Q$ are respectively given by,
\begin{equation}
	\{Q\}_\text{even} = \frac12(Q+\beta Q\beta),\quad \{Q\}_\text{odd} = \frac12(Q-\beta Q\beta)~.
\end{equation}

As $\beta$ is an even operator and $J$ is an odd operator, Eq.~(\ref{UHU}) is an even expression which does not mix the positive and negative energy states. The EFW transformation has the further benefit that in many cases the odd components of $\sqrt{H^2}$ vanishes. In practice $\sqrt{H^2}$ is taken as a perturbative expansion where the rest mass energy is the dominate term.  

In Ref.~\cite{goncalves07}, Goncalves \textit{et al.} report to use the EFW transformation to arrive at the following non-relativistic limit of Eq.~(\ref{H_GW}):
\begin{equation}
\begin{split}
H_\text{GOS}=&\frac{1}{2m}(\delta^{ij}+2fT^{ij})[(p_i-eA_i)(p_j-eA_j)\\
	&+e\hbar\epsilon_{jkl}\sigma^l\partial^k(A_i)]\\
	&+\frac{\hbar}{2m}\partial^i(f)T^{jl}\epsilon_{ijk}\sigma^k(p_l-eA_l)+mc^2~,
\end{split}
\label{H_GOS}
\end{equation}
where $T\equiv\text{diag}(0,-1,1)$ and partial derivatives act only on the contents of the parenthesis which follow. $\delta^{ij}$, $\epsilon^{ijk}$, and $\sigma^i$ are the Kronecker delta, Levi-Civita symbol, and Pauli matrices respectively. 

The spin equation of motion is given by $i\hbar d\sigma_i/dt=[\sigma_i,H]$. Calculating the commutation with $H_\text{GOS}$ and then taking the $\hbar\rightarrow0$ limit, one arrives at the following semi-classical spin equation of motion,
\begin{equation}
\begin{split}
\frac{d\sigma_i}{dt}=&\frac{e}{m}\epsilon_{ijk}\sigma^k\epsilon^{jlm}\partial_l(A_m+2fT_{mn}A^n)\\
&-\frac{1}{m}(p^j-eA^j)\sigma^k[T_{kj}\partial_i(f)-T_{ij}\partial_k(f)]~.
\end{split}
\label{spin_EFW}
\end{equation}

From Eq.~(\ref{spin_EFW}), Goncalves \textit{et al}. claims that even in the absence of a magnetic field, spin precession may still occur due to the coupling of the GW to the EM gauge field. Setting the magnetic field $B^i=\epsilon^{ijk}\partial_j(A_k)=0$ and neglecting small $\partial(f)$ (which is an appropriate approximation for GWs of astronomical sources on Earth) the spin precession is, 
\begin{equation}
\frac{d\sigma_i}{dt}=\frac{2e}{m}f\epsilon_{ijk}\epsilon^{jlm}T_{mn}\partial_l (A^n)\sigma^k~.
\label{spin_EFW_short}
\end{equation}

This poses the notion that due to the presence of the GW, the gauge field can have a physical effect on spin precession even in the absence of a magnetic field. We argue that the contribution to spin precession as presented by Eq.~(\ref{spin_EFW}) and (\ref{spin_EFW_short}) cannot be physical, as it is not gauge invariant. More generally, the GW correction term to the magnetic dipole energy in $H_\mathrm{GOS}$, i.e. the term proportional to $T^{ij}\epsilon_{jkl}\sigma^l\partial^k(A_i)$, is gauge-variant under the usual gauge transformation, $A_i\rightarrow A_i+\partial_i \chi$ and $\psi\rightarrow e^{i\frac{e}{\hbar}\chi}\psi$, where $\chi$ is some scalar function. We believe the gauge-variant term in $H_\text{GOS}$ is not the result of the EFW transformation, but the result of an erroneous calculation in the application of the EFW transformation, as our derivation using the EFW transformation yields no such gauge-variant term.

We begin by writing Eq.~(\ref{H_GW}) in a more convenient form,
\begin{equation}
	H=\beta mc^2+c\alpha^j(\delta_j^i+T_j^if)(p_i-eA_i)~.
\end{equation}

Hence,
\begin{equation}
\begin{split}
	H^2=&~m^2c^4+c^2\alpha^j\alpha^k(\delta_j^i+T_j^if)\\
		&\times [(\delta_k^l+T_k^lf)(p_i-eA_i)(p_l-eA_l)\\
		&-T_k^lp_i(f)(p_l-eA_l)].
\end{split}
\end{equation}

Neglecting the small $v^2$ order terms, and using the identity $\alpha^i\alpha^{j}=i\epsilon^{ijk}\sigma_k \textbf{I}_{2} + \delta^{ij}\textbf{I}_{4}$,
\begin{equation}
\begin{split}
	H^2=&~m^2c^4+c^2(\delta^{ij}+2fT^{ij})[(p_i-eA_i)(p_j-eA_j)]\\
	&+\frac{e\hbar c^2}{2}(\delta^{ij}+2fT^{ij})\epsilon_{jkl}\Sigma^l[\partial^k(A_i)-\partial_i(A^k)]\\
	&+\hbar c^2\partial^i(f)T^{jl}\epsilon_{ijk}\Sigma^k(p_l-eA_l)~.
\end{split}
\end{equation}
where $\Sigma_{k}\equiv \sigma_k \textbf{I}_2$. As the rest mass energy is the dominate energy term in the non-relativistic limit the perturbative expansion of $\sqrt{H^2}$ to $O[1/m^2]$ accuracy yields,
\begin{equation}
\begin{split}
H_\text{EFW}=&\frac{1}{2m}(\delta^{ij}+2fT^{ij})[(p_i-eA_i)(p_j-eA_j)]\\
&+\frac{e\hbar}{4m}(\delta^{ij}+2fT^{ij})\epsilon_{jkl}\sigma^l[\partial^k(A_i)-\partial_i(A^k)]\\
&+\frac{\hbar}{2m}\partial^i(f)T^{jl}\epsilon_{ijk}\sigma^k(p_l-eA_l)+mc^2~.
\end{split}
\label{H_EFW}
\end{equation}

The first and second terms of $H_\text{EFW}$ involves the kinetic and magnetic dipole energies and their corrections due to the GW. The third term can be thought of as being the GW analogue of the Schwarzschild gravitational spin-orbit energy~\cite{oliveira62}. The last term is the rest mass energy. In comparison to the $H_\text{GOS}$, $H_\text{EFW}$ has an extra term proportional to $\partial_i(A^k)$ in the second line of Eq.~(\ref{H_EFW}). The presence of this term has the effect of ensuring gauge-invariance in the Hamiltonian. 

One notes that the EFW transformation has been known to produce spurious parity-violating terms~\cite{nicolaevici02,obukhov02a,silenko05,jentschura14}. Ref.~\cite{jentschura14} argues that these parity-violating terms are the result of chiral transformation of the $U_2$ operator, which alters the symmetry properties of the Hamiltonian. We point out that the gauge-variant term in $H_\text{GOS}$ is not of this type. 

To further our claim that $H_\text{EFW}$ is the correct non-relativistic limit of the Dirac Hamiltonian in a GW background, we provide a consistency check by arriving at the same Hamiltonian using the alternative SFW transformation. 

\section{Standard Foldy-Wouthuysen Transformation}
\label{sec:sfw}

The odd and even components of $H$ are respectively given by,
\begin{equation}
O=\frac12(H-\beta H\beta),\quad E=\frac12(H+\beta H\beta)~.
\label{odd_even}
\end{equation}

The SFW is a multi-commutator expansion of the unitary transformation $U=e^{iS}$,
\begin{equation}
H'=H+i[S,H]+\frac{i^2}{2!}[S,[S,H]]+\cdots~,
\label{H_prime}
\end{equation}
where $S=-\frac{i\beta}{2m}O$. $i[S,H]\approx-O$ generates a term that eliminates the odd operator $O$, however many more terms are generated by the higher-order terms which potentially could be odd operators. To eliminate these odd operators, the FW transformation is repeated on subsequent Hamiltonians (i.e. $H',H'', H'''$, and so on) until all odd operators are eliminated to the required order of accuracy. 

For convenience we will work in the natural units where $\hbar=c=e=1$. We will put $\hbar,c,e$ back into the final equation. We begin by calculating the following commutator relations:
\begin{align}
&[\beta\alpha^i(p_i-A_i),\alpha^j(p_j-A_j)]\nonumber\\
&	\quad=2\beta\delta^{ij}(p_i-A_i)(p_j-A_j)+2\beta\epsilon^{ijk}\Sigma_k\partial_j(A_i)~,\\
&[\beta\alpha^i(p_i-A_i),f\alpha^2(p_2-A_2)]\nonumber\\
&	\quad=2\beta f(p_2-A_2)^2+\beta f\epsilon^{i2k}\Sigma_k[\partial_2(A_i)-\partial_i(A_2)]\nonumber\\
&	\quad\quad+\beta\Sigma_3\partial_1(f)(p_2-A_2)~,\\
&[\beta\alpha^i(p_i-A_i),f\alpha^3(p_3-A_3)]\nonumber\\
&	\quad=2\beta f(p_3-A_3)^2+\beta f\epsilon^{i3k}\Sigma_k[\partial_3(A_i)-\partial_i(A_3)]\nonumber\\
&	\quad\quad+\beta\Sigma_2\partial_1(f)(p_3-A_3)~.
\end{align}

Using these commutator relations and Eq.~(\ref{H_GW}) we write down,
\begin{align}
&i[S,H]=\frac{\beta}{m}\delta^{ij}(p_i-A_i)(p_j-A_j)+\frac{\beta}{m}\epsilon^{ijk}\Sigma_k\partial_j(A_i)\nonumber\\
&	\quad+\frac{2\beta}{m} fT^{ij}(p_i-A_i)(p_j-A_j)\nonumber\\
&	\quad+\frac{\beta}{m} fT^{ij}\epsilon_{jkl}\Sigma^l[\partial^k(A_i)-\partial_i(A^k)]\nonumber\\
&	\quad+\frac{\beta}{m}\partial^i(f)T^{jl}\epsilon_{ijk}\Sigma^k(p_l-A_l)\nonumber\\
&	\quad-\alpha^i(p_i-A_i)+f[\alpha^2(p_2-A_2)-\alpha^3(p_3-A_3)] + \text{h.o.}
\label{iSH}
\end{align}
where h.o. indicate that there are higher order terms.

Using Eq.~(\ref{iSH}),
\begin{align}
&i[S,H]+\frac{i^2}{2}[S,[S,H]]=\frac{1}{2m}[\beta\alpha^i(p_i-A_i),\nonumber\\
&\quad\frac12\alpha^j(p_j-A_j)-f\alpha^2(p_2-A_2)+f\alpha^3(p_3-A_3)+\beta m]\nonumber\\
&\quad+\frac{1}{2m}[-\beta f\alpha^2(p_2-A_2)+\beta f\alpha^3(p_3-A_3),\beta m] + \text{h.o.}
\end{align}

Neglecting higher order terms,
\begin{align}
&\mathcal{H}_\text{SFW}=H + i[S,H]+\frac{i^2}{2}[S,[S,H]]\nonumber\\
&\quad= \beta m + \frac{\beta}{2m}(\delta^{ij}+2fT^{ij})(p_i-A_i)(p_j-A_j)\nonumber\\
&\quad+\frac{\beta}{4m}(\delta^{ij}+2fT^{ij})\epsilon_{jkl}\Sigma^l[\partial^k(A_i)-\partial_i(A^k)]\nonumber\\
&\quad+\frac{\beta}{2m}\partial^i(f)T^{jl}\epsilon_{ijk}\Sigma^k(p_l-A_l)~.
\label{H_SFW_planck}
\end{align}

Explicitly reinstating $\hbar,c,e$, one retrieves Eq.~(\ref{H_EFW}) from Eq.~(\ref{H_SFW_planck}). In other words $H_\text{EFW}=H_\text{SFW}$ to $O[1/m^2]$ accuracy, where $\mathcal{H}_\text{SFW}=\beta H_\text{SFW}$. The EFW and SFW transformations are not equivalent unitary transformations, and in general can give rise to different Hamiltonians in the non-relativistic limits. However in the current case, that both the EFW and SFW transformation yields the same Hamiltonian, is good verification that Eq.~(\ref{H_EFW}) is the correct non-relativistic limit of the Dirac Hamiltonian in the presence of an EM gauge and GW field. 

Importantly, $H_\text{EFW}$ is gauge invariant. This has consequences for the physical behaviour of spin-1/2 particle interacting with GWs. In particular there is stark difference in the spin-precession behaviour described by $H_\text{GOS}$ and $H_\text{EFW}$. The corresponding semi-classical spin equation of motion under $H_\text{EFW}$ is 
\begin{equation}
\begin{split}
	\frac{d\sigma_i}{dt}=&\frac{e}{m}\epsilon_{ijk}\sigma^k\epsilon^{jlm}[\partial_l(A_m)+fT_{mn}\partial_l(A^n)-fT_{mn}\partial^n(A_l)]\\
		&-\frac{1}{m}(p^j-eA^j)\sigma^k[T_{kj}\partial_i(f)-T_{ij}\partial_k(f)]~.
\end{split}
\label{spin_SFW}
\end{equation}
Unlike the spin equation of motion derived from $H_\text{GOS}$, Eq.~(\ref{spin_SFW}) is gauge-invariant. If one neglects the small $\partial(f)$ and set the magnetic field to zero then $d\boldsymbol{\sigma}/dt=0$, and there is no spin precession, in contrast to Eq.~(\ref{spin_EFW_short}) and the claims of Goncalves \textit{et al.}

\section{Conclusion}
The spin-precession of a spin-1/2 particle in the absence of a magnetic field but in the presence of a GW is a false effect of a gauge-variant Hamiltonian. We consistently derived a gauge-invariant Hamiltonian with both the EFW and SFW transformations, which we contend to be the correct non-relativistic limit of the generalised Dirac Hamiltonian in the background of a GW spacetime metric. Using this Hamiltonian we showed the novel spin-precession effect no longer exists.

\section*{Acknowledgements}

The author would like to thank M. Lajk\'{o}, C.-H. Su, A. Martin, and S. Quach for discussions and checking the manuscript. This work was financially supported by the Japan Society for the Promotion of Science. The author is an International Research Fellow of the Japan Society for the Promotion of Science.

\bibliography{gwfw}

\end{document}